# Liquid Mirror Telescopes: A Review

Ermanno F. Borra

Centre d'Optique, Photonique et Laser, Département de Physique, Université Laval,
Québec, Qc, Canada G1K7P4
http://wood.phy.ulaval.ca/lmt/home.html
Fax: 418-656-2040: Email: borra@phy.ulaval.ca

## ABSTRACT

I review the present status of liquid mirror telescopes. In a nutshell: LMTs do work and several have been built and used. Extensive interferometric tests of liquid mirrors (the largest one having a diameter of 2.5 meters ) show diffraction limited performance. A large quantity of astronomical data have been taken at the NASA 3-m LMT and are currently being analyzed. Finally, I shall address the future of the technology as well as the issues of the field accessible to LMTs equipped with novel optical correctors.

## 1. INTRODUCTION

The surface of a spinning liquid takes the shape of a paraboloid that can be used as the primary mirror of a telescope. Following the suggestion (Borra 1982) that modern technology gives us tracking techniques that render liquid mirrors useful to Astronomy, a research and development program was begun to assess the feasibility of the concept, leading to the demonstration of a diffraction limited 1.5-m mirror (Borra et al. 1992), an article that also gives a wealth of technological details. This was followed by a 2.5-m mirror which was also extensively tested ( Borra, Content, & Girard, 1993; Girard, & Borra, 1997). Liquid mirror telescopes cannot be tilted and hence cannot track like conventional telescopes do. To track with imagery, narrow-band filter spectrophotometry or slitless spectroscopy, one can use a technique, called time delayed integration (TDI), that uses a CCD detector that tracks by electronically stepping its pixels. The information is stored on disk and the nightly observations can be coadded with a computer to give long integration times. The technique has been demonstrated (Hickson et al. 1994) with a 2.7-m diameter liquid mirror telescope. They show a deep sky exposure taken with that telescope.

Why should one be interested in liquid mirror telescopes, considering their limitations? The main reason comes from the size and cost advantages. The low cost makes it possible for a small team of astronomers to have their own large telescope working full-time on a specific project. This is in practice not realistic with expensive classical telescopes. Some research projects (e.g. time consuming surveys) simply cannot be envisioned with classical telescopes but are possible with LMTs. This is particularly true for the types of research where the region of sky observed is not particularly important (e.g. cosmology).

The issue of the field of regard accessible to a liquid mirror is an important one, for the usefulness of a LMT increases greatly with it. We have begun investigating this topic, finding a corrector design that gives a very large field of regard.

At the time of this writing several liquid mirrors and liquid mirror telescopes have been built. This review reports on the present status of liquid mirrors and liquid mirror telescopes. It ends with a recommendation as to what telescope should be built NOW.

## 2. LIQUID MIRRORS

The surface of a reflecting rotating liquid takes the shape of a parabola which is the ideal surface for the primary mirror of an astronomical telescope. The focal length of the mirror L is related to the acceleration of gravity g and the angular velocity of the turntable ω by

$$L = g/(2\omega^2) \tag{1}$$

For large mirrors of practical interest the periods of rotation are of the order of several seconds and the linear velocities at the rims of the mirrors range between 2 and 10 km/h. Figure 1 shows an exploded view of the basic mirror setup.

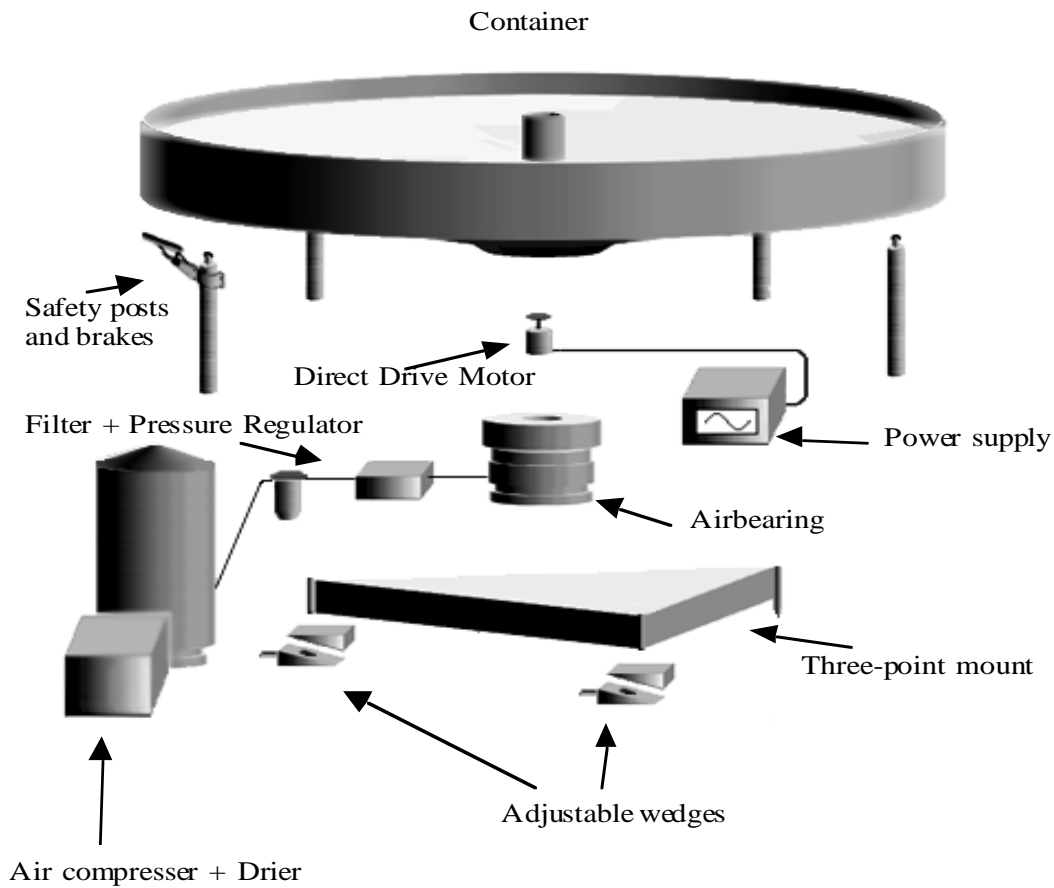

Figure 1. It shows an exploded view of the basic mirror setup.

The  container and bearing rest on a three-point mount that aligns the axis of rotation parallel to the gravitational field of the Earth. Ideally the alignment should be done to better than 0.25 arcseconds. Larger alignment errors give acceptable images but introduce scattered light (Girard & Borra 1997).

All existing liquid mirrors use airbearings because they are convenient for small systems and commercially available units have the required precision and low friction. Oil-lubricated bearings should be more advantageous for large mirrors since they  can support substantially higher masses for a given bearing size. We are presently experimenting with a ball bearing but do not have quantitative test results at the time of this writing. The turntable is driven by a synchronous motor which is controlled by a variable-frequency AC power supply stabilized with a crystal oscillator. The container is a very important component of the system. It must be light, and rigid. We currently make them with Kevlar laminated over a foam core (Hickson, Gibson & Hogg, 1993). Other construction techniques can be used. It may be advantageous to use space frames for very large mirrors. We  have developed techniques that allow us to work with layers of mercury as thin as 0.5 mm (Borra,  Content, Girard, Szapiel,  Tremblay, & Boily 1992). On cannot overstress the importance of working with thin mercury layers to minimize weight and, especially, to dampen disturbances.

The construction of the 3.7-m mirror is documented on the Laval LMT  Web site: http://wood.phy.ulaval.ca/lmt/home.html. Table 1 shows an estimate of the cost of building this mirror. This is a prototype and a better engineered system would certainly be less expensive. It must also be noted that these are rough estimates. On the one hand there are hidden costs that are not accounted for (e.g. rent for construction space), on the other the construction time can be decreased given our experience building the present containers.

Table 1.
Costs of components and labor needed to construct a  3.7-m mirror. Adapted from Girard (1997).

| Item | Cost of components (1996 US $) | Labor ($45/hour) |
|---|---|---|
| Mirror + accessories[1] | 32,000 | 22,000 |
| Safety equipment[2] | 4,000 | 6,00 |
| Installation[3] | 1,000 | 4,000 |
| Total | 37,000 | 32,000 |

[1] Complete system, including base, mercury, motor, etc...
[2] Includes mercury sniffer, safety brakes and anti-spill wheels
[3] In-situ installation, including balancing, debugging, checking image quality .

We have extensively tested a 2.5-m f/1.2 LM (Borra, Content & Girard 1993; Girard & Borra 1997). We use null lenses to correct the large spherical aberration present at the center of curvature of any parabolic mirror. The interferometry is done with scatter plate and Shack-cube interferometers. The interferograms are captured with 1/60 second exposure times with a CCD detector  and computer analyzed.

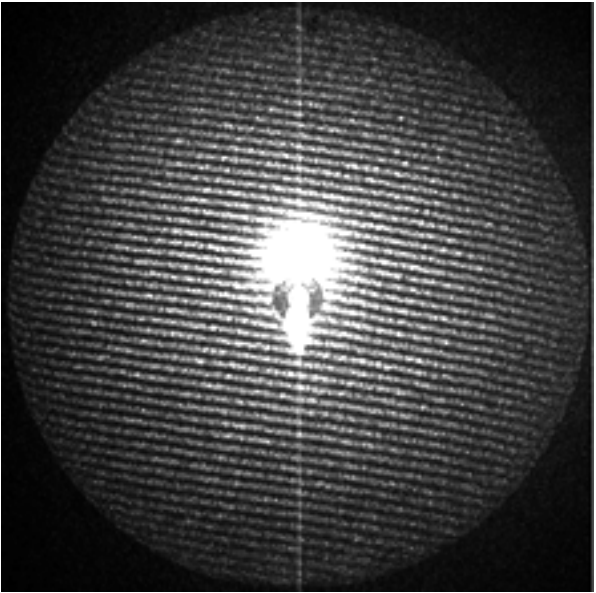

Figure 2. It shows a typical interferogram of the 2.5-m mirror.

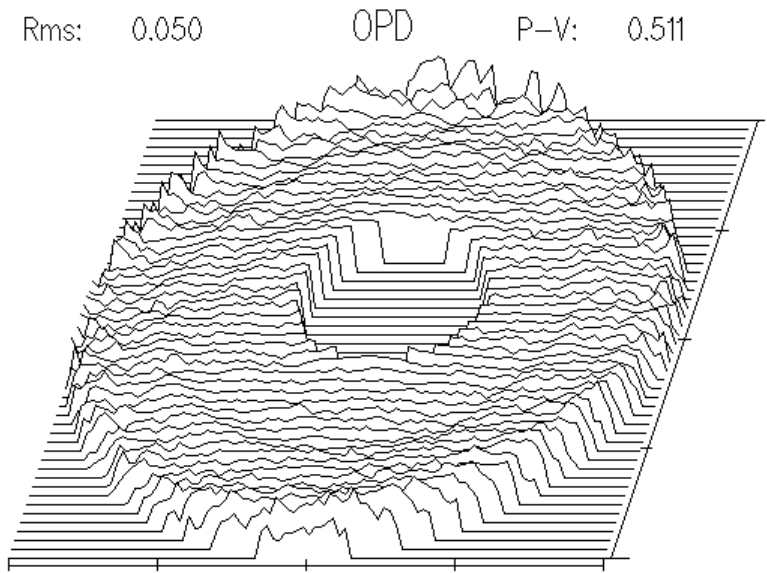

Figure 3. It shows the three-dimensional rendering of the surface of the 2.5-m mirror.

Figure 2 shows a typical interferogram of the 2.5-m mirror and Figure 3 shows the three-dimensional rendering of its surface. The statistics associated with the surface are given in units of surface deviations on the mirror at a wavelength of 6328 Å. The spatial resolution of the interferometry on the mirror is 3 cm. The 1/60 second capture times are sufficiently short that we can detect rapid liquid movements, but they also render the interferometry sensitive to seeing effects in the testing tower. Our testing facility is in a basement room and has little local seeing. Because the mirror is liquid, a few selected interferograms are not necessarily representative of its optical quality. We have analyzed a large number of interferograms and have videotaped hours of interferogram data to satisfy ourselves that the interferogram shown in Figure 2, and the wavefront of Figure 3 are representative. The statistics of several sequences of wavefronts have been published in several places (e.g. Borra, Content , & Girard 1993). They can be summarized by stating that the root mean square values of the deviations from a perfect parabola are typically $\lambda/20$ for a well-tuned mirror. Good tuning involves a stable rotational velocity and an axis of rotation aligned within a fraction of one arcsecond from the vertical..

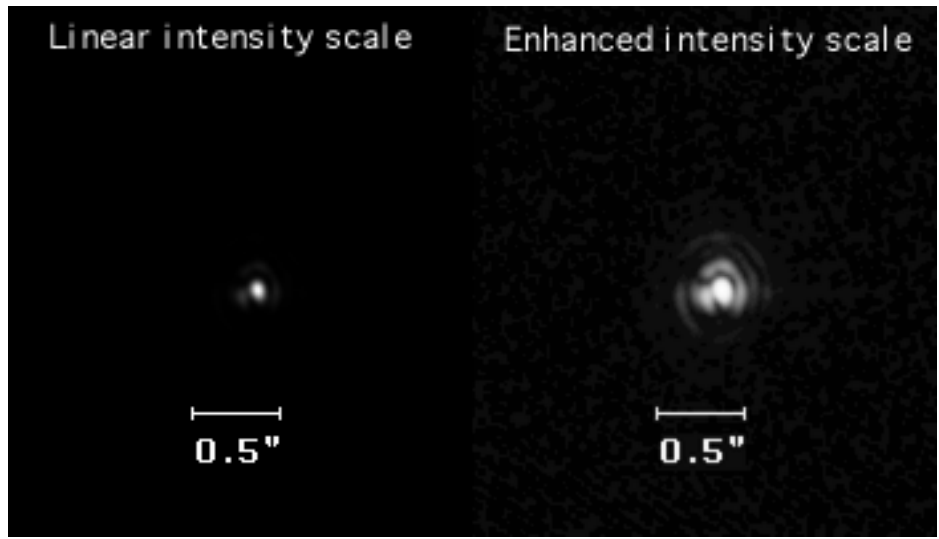

Figure 4. It shows the image of an artificial star imaged, with the 2.5-m mirror.

Figure 4 shows the image of an artificial star imaged, through null lenses, at the center of curvature of the 2.5-m mirror. It shows the Airy diffraction pattern of the 2.5-meter diameter liquid mirror. Fainter rings are present further away from the center of the PSF but are below detection in this image. We have videotaped hours of data and find that the Airy pattern is always visible, although the intensities and symmetries of the rings vary a little, some contribution coming from seeing in the testing tower. The observation of the Airy pattern (taken through the null lenses) gives direct evidence that the mirror is near diffraction limit.

Liquid mirrors are sensitive to vibrations but these are only a minor nuisance. Our mirrors are located in the basement of a large building that vibrates like any building does. We do see the effect of vibrations in the form of concentric rings on the surface of the mirror. However, their amplitudes are negligible ( ~1/100 $\lambda$) for thin mercury layers. We can walk around our mirrors without inducing excessive vibrations.

We have made detailed analysis of the scattered light of liquid mirrors. We find that liquid mirrors do not have an undue amount of scattered light. To minimize scattered light, one must work with a thin layer of mercury to dampen vibration and wind effects. We also found that misalignments of the axis of rotation of the mirror introduce scattered light. Ideally, the axis should be vertical within 0.25 arcseconds; although errors below 1 arcseconds introduce tolerable levels of scattered light. The behavior of the 2.5-m mirror under perturbation has been extensively studied (Girard 1997, Girard & Borra 1997). The telescope should also be enclosed in a plastic shroud to protect it from the wind.

## 3. TELESCOPE AND OBSERVATORY

Figure 5 shows the entire telescope system. Comparing the LMT to a conventional telescope, we see that they are similar with the exception of the mount. The top parts are identical, consisting of a focusing system and a detector: there is some cost saving in the upper end structure since it does not have to be tilted. The largest cost savings obviously come from the mount which consists of a simple tripod.

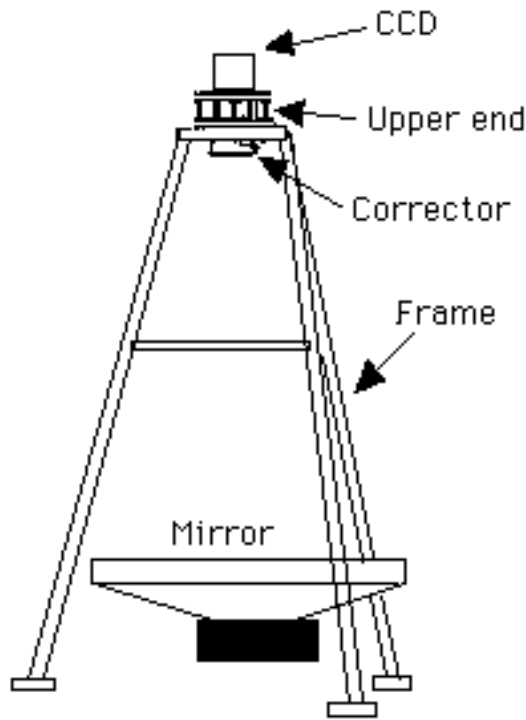

Figure 5: It shows the entire telescope system.

Figure 6 shows a layout of the telescope and observatory for a 4-m telescope. The structure is much simpler than the dome of a conventional telescope. The roof and the folding platform needed to service the upper end are the only movable parts and are inexpensive. The structure consists of a steel frame with metal sidings. Table 2 gives a cost estimate breakdown of the entire system. Shipping, and installation costs as well as the computing facilities are not included. On the basis of experience with the NASA 3-m telescope, one can estimate that it will take 6 months and a full time operator to debug the system. After the debugging period, the ratio of maintenance/running time will be about 1/5. After a while it should decrease to 1/10. At present, all running LMTs have used a full time operator but there are no compelling reasons why an operator should be present at all times at the telescope. LMTs could be fully robotics and only routine maintenance would be needed. The present experience with LMTs indicate that yearly operation costs should be about 10% of capital costs.

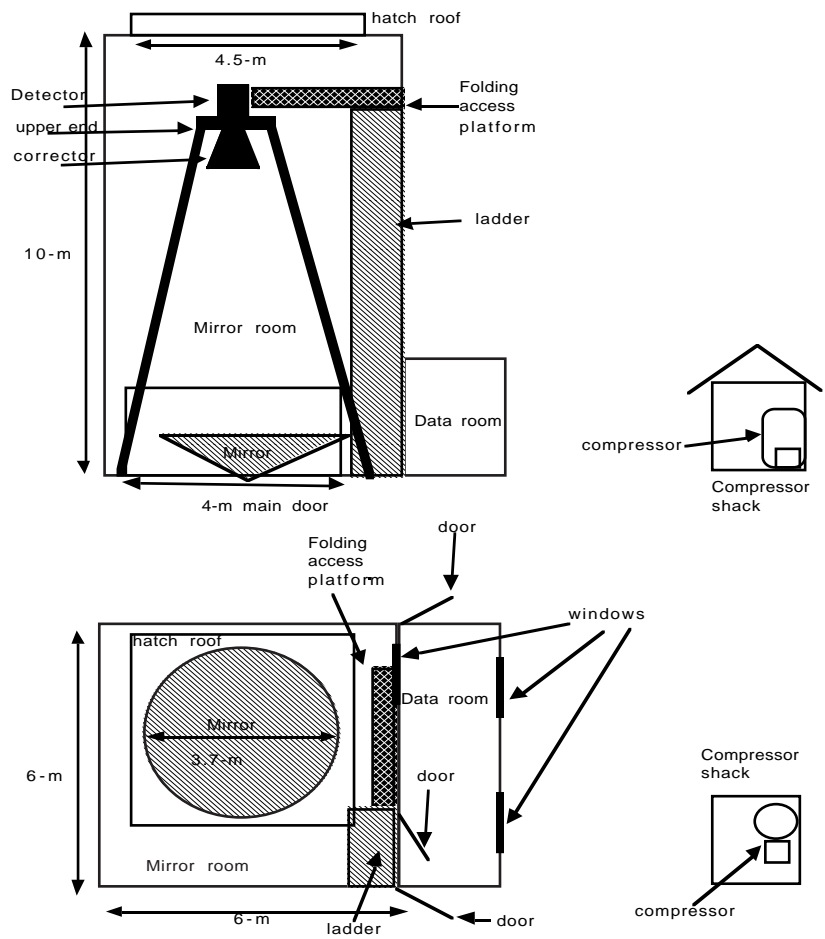

Figure 6: It shows a layout of the telescope and observatory for a 4-m telescope.

Table 2
Cost breakdown of telescope and observatory

========================================================

| Item | Cost (1000 US$) | based on |
| --- | --- | --- |
| Building: | 75 | Architect + quotes |
| Mirror: | 60 | UL 3.7-m |
| Upper end: | 110 | CFHT IR upper end |
| Corrector: | 35 | Quote for 5-m(56K$) |
| CCD: | 75 | G. Walker |
| Salaries: | 80 | PDF 2 years |

### 4. CURRENT APPLICATIONS OF LMTs.

At the time of this writing several LMTs have been built and operated for scientific applications. The earliest observations were taken with a makeshift observatory housing a 1.4-m liquid mirror. Although really meant to be a demonstration more than a serious scientific project, they did yield published astronomical research (Content et al. 1989). The first professional-quality effort was done with a 2.7-m LMT used at the University of Western Ontario for atmospheric physics that has been in operation for a few years (Sica et al. 1995, Sica & Thorsley 1996). The first professional-quality astronomical telescope was a 2.7-m LMT built and operated at the University of British Columbia. It demonstrated that LMTs can obtain deep astronomical images with resolution limited only by atmospheric seeing (Hickson et al. 1994). Poor weather limited the scientific output of the instrument. P. Hickson is now replacing it by a larger and more advanced LMT (See de Lapparent's contribution in these proceedings). NASA has built and operated a very successful 3-m imager that has yielded a large number of astronomical observations. A pretty image obtained by Mark Mulrooney with that telescope is shown in Figure 7.

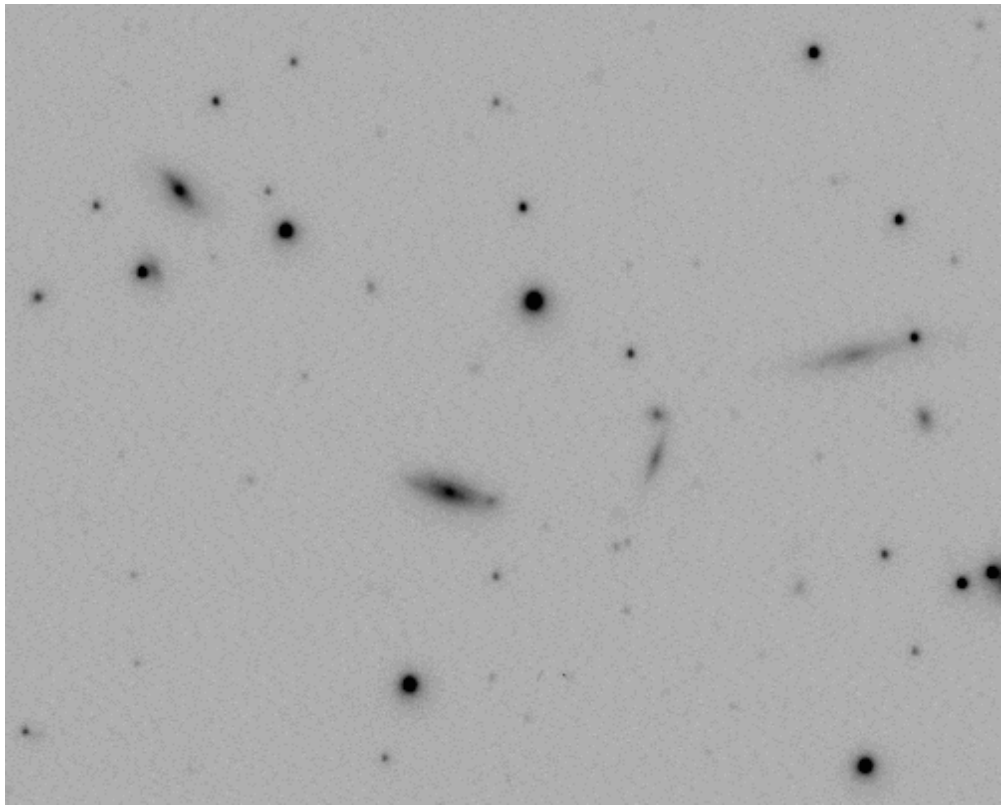

Figure 7. It shows a CCD image of a region of sky 5 arcminutes long observed with a 3-m liquid mirror telescope built and operated by NASA. The image is a courtesy of Mark Mulrooney.

It was obtained from a single nightly pass that gives a 100-second exposure. It reaches about magnitude 23 and shows several faint galaxies and stars. Cabanac discusses 3-m data in greater details in these proceeding. UCLA has recently built a lidar facility, similar to the UWO system, in Alaska. Liquid mirrors are also used to generate reference surfaces to test optical surfaces. The work of Ninane & Jamar (1996), is particularly of interest, not only because of their original use of a liquid mirror, but also because they carried out independently optical shop tests that confirm the high optical quality of liquid mirrors reported by the Laval team. It also worthwhile to point out an interesting application of active flat mercury mirrors (Ragazzoni & Marchetti 1994).

## 5. FIELD CORRECTORS FOR LIQUID MIRROR TELESCOPES

The outstanding limitation of liquid mirror telescopes comes from the fact that they cannot be tilted and therefore are limited to zenith observing. Because the aberrations of a parabola increase rapidly with field angle, classical corrector designs cannot yield subarcsecond images for angles significantly greater than one degree. To access larger fields, innovative corrector designs must be explored. Borra (1993) has explored analytically how far off-axis one can use a liquid mirror telescope and has shown that, in principle, aberrations can be corrected for zenith distances as large as 45 degrees.

A first exploration Borra, Moretto and Wang (1995), investigated a two-mirror corrector used with a parabolic primary mirror observing at large zenith angles. The design, dubbed BMW corrector, uses complex surfaces that give subarcsecond images as far as 22.5 degrees from the optical axis. However, the required complex anamorphic aspheric surfaces cannot be made economically with existing technology .

As a further step toward a design feasible with existing technology, Moretto and Borra (1996) investigated the BMW design assuming mirrors that can be made with the Active Vase Mirror technology pioneered by Lemaître (1981, 1989). Laboratory tests (Moretto et al. 1995) show that it is possible to generate the complex figures needed by mechanically stressing a steel mirror. The BMW corrector is further discussed by Moretto, Lemaître and Ferrari in these proceedings.

Once a sufficiently large accessible field is achieved, a fixed primary and movable correctors yield a more efficient system than a classical tiltable telescope. A classical telescope can only observe a field at a time, while a fixed primary with several correctors could access many widely separated fields simultaneously. This primary-sharing setup, allowing several research programs to be carried out simultaneously, is particularly attractive for very large telescopes.

## 6. FUTURE DEVELOPMENTS

This is a very young technology that is bound to undergo considerable improvements. I briefly try to predict the major future developments.

1) How large? One of the most often asked questions concerns the maximum size that LMTs may eventually reach. The wind generated by the rotation of the mirror is the single problem most likely to set a maximum diameter to the size of LMs. We do not have quantitative estimates of this limit but, on the basis of the results with the present LMs, I would estimate that a 6-m to 8-m mercury mirror is probably feasible. Larger sizes are possible, especially with other reflecting liquids. In practice, considering detector matching considerations, it is not clear that it is worthwhile to build a single 30-m LMT. It is probably better to build an array of 6-m or 8-m LMTs to reach much larger effective collecting areas (30 to 100-m). Note that costs are likely to decrease with technological improvements.

2) Reflecting liquids: We are now using mercury because it is easy to handle. Note that the toxicity of mercury vapors is, in practice, not a significant problem (Borra et al. 1992). We have begun to experiment with gallium and gallium alloys, with encouraging results. Note that they are very easy to supercool and stable in that sate. We have supercooled them to -30 degrees C. and have operated a telescope to subfreezing temperatures (Borra et al. 1997). We have begun experimenting with other reflecting liquids that have low densities and high viscosities, two important characteristics for very large inexpensive LMs.

3) Correctors for LMTs: The field of regard of LMTs can certainly be increased. The BMW corrector is very promising and likely to deliver 20-degree fields of regard. Off-axis correctors are an unexplored field of research in optical design and improved design could do better. It is no inconceivable that 90 degree fields may be obtained one day, considering that there are no fundamental reasons to prevent this (Borra 1993). Holographic correctors are an interesting possibility. We did explore the design a few years ago (Lemelin, Lessard and Borra 1993) concluding that the technology was not ready. However, spurred by commercial applications, research in holographic optical elements is progressing at a furious pace and holographic correctors may soon become practical.

4) Instrumentation for LMTs: Although only imagery has been demonstrated so far with LMTs, I have argued (Borra 1987) that essentially all types of astronomical instrumentation could be adapted to LMTs. Dohlen discusses spectroscopy for LMTs in these proceedings. Recently, Texas Instruments has introduced a chip that contains a large number of tiny tiltable mirrors. Placed in the focal plane, they could be used to define small regions that deflect light into a spectrograph. This opens the way for making the equivalent of movable apertures in the focal plane, allowing thus to perform aperture spectroscopy with an LMT. A new generation of three-dimensional photon counting imagers that measures the energy of the photon is being developed. This open the way for imaging spectrophotometry.

# 7. WHAT CAN BE DONE NOW?

The issues to address are: If we want to build an LMT now, what telescope size should we build; with what instrumentation; with what corrector?

1) Mirror size

At the time of this writing, a 4-m class LMT is a low-risk undertaking, albeit it will still be a pioneering effort. We can say so because it is a small extrapolation from the 3-m NASA LMT and the 2.5-m LMT that has been tested in our laboratory. We can use the same technology . Content's finite element computations (these proceedings) show that a composite container, like those built so far, should do the job.

2) Instrumentation

Tracking should be done with the TDI technique, which clearly works. This restricts us to direct imagery. Obviously low-resolution spectroscopy can be carried out with interference filters. Spectroscopy with an LMT has not been demonstrated yet but it can be done (Dohlen in these proceedings). It must be noted that, with a classical corrector, the TDI technique degrades the images. This comes from the fact that the TDI technique moves the pixels on the CCD at a constant speed on a straight line while the images in the sky move at different speeds on curved trajectories (Gibson & Hickson, 1992) . The deformation depends on the latitude of the observatory (it is zero at the equator and increases with latitude). However, optical design shows that it is possible to eliminate the effect by introducing the adverse deformation on the field with the corrector.

3)Corrector

The corrector of choice consists of a semi-classical on-axis glass corrector capable of about 1-degree field. It should remove the TDI distortion. A slightly off-axis corrector (1.5 degrees) is an intriguing possibility since it would be possible to cover a 3-degree strip of sky by moving it from -1.5 degrees to +1.5-degree. Harvey Richardson (private communication) has made such design. The BMW corrector discussed in these proceedings is probably  feasible but it is too experimental. Also, it should essentially double the cost of the observatory.

## 8.  CONCLUSION

Interferometric tests of liquid mirrors having diameters as large as  2.5 meters show excellent optical qualities.  Although  there is room for improvement of this very young technology,  we do have a working design that is sufficiently robust to be useful for practical use. We also have an adequate understanding of the behavior of a liquid mirror under perturbations. In other words, we know how to make liquid mirrors that work but one can do better.

A handful of liquid mirrors have now been built and are used for scientific work: astronomy, atmospheric sciences, space sciences, optical shop tests. More applications are certainly forthcoming given the advantages of liquid mirrors, foremost of which is cost.

The issue of the field accessible to a LMT is a very important one since the usefulness of a LMT increases with its accessible field. Optical design work indicates that a single LMT should be able to access fields as large as 45 degrees. We only are at the beginning of this exploratory work and additional effort can probably find simpler systems with improved performance. On the other hand, those correctors can only increase the cost of a LMT and they can only be practical if the total instrument is sufficiently cheaper than a conventional telescope.

The present state of the technology is such that an astronomical observatory housing a 4-m class LMT is a low-risk undertaking.

## 9.  ACKNOWLEDGMENTS

This research has been supported by Natural Sciences and Engineering Research Council of Canada and Formation des Chercheurs et Aide à la Recherche grants. A very large number of people have contributed to the success of Liquid Mirrors. Some of them, but not all, are named in the reference list. I thank them all.

## 10.  REFERENCES

Borra, E.F.  1982, JRASC 76, 245.
Borra, E.F. 1987,  PASP 99, 1229.
Borra, E. F., 1993, A&A 278,665.
Borra,  E.F.,  Content, R.,. Girard, L, Szapiel, S.,  Tremblay, L.M., & Boily, E.  1992, ApJ 393, 829.

Borra, E.F., Content, R., & Girard, L 1993, ApJ 418, 943.

Borra, E.F, Moretto, G., & Wang, M. 1995, A&A 109, 563.

Borra, E.F., Tremblay, G., Huot , Y., & Gauvin, J. 1997, PASP, In Press.

Content, R., Borra, , E.F. Drinkwater, M.J., Poirier, S., Poisson, E., Beauchemin, M., Boily, E., Gauthier, A., & Tremblay, L.M 1989, AJ, 97, 917.

Girard, L. 1997, Ph.D. thesis, Université Laval.

Girard, L & Borra E.F. 1997, applied Optics In Press.

Gibson, B.K., & Hickson, P. 1992, MNRAS 258, 543.

Hickson, P., Borra, E.F., Cabanac, R. , Content, R., Gibson, B.K, & Walker, G. A. H. 1994, ApJ 436, L201.

Hickson, P., Gibson, B.K., & Hogg, D.W. 1993, PASP, 105, 501.

Lemaitre, G. 1981, in Current Trends in Optics, International Commission for Optics, Taylor & Francis Publ., London, 135.

Lemaître, G. 1989, Various aspects of active optics, in: Proc. SPIE on Telescopes and Active Systems, Orlando, FA, 328.

Lemelin, G., Lessard, R. and Borra, E.F., 1993, A&A 274, 983.

Moretto, G., & Borra , E.F. 1996, Applied Optics In Press.

Moretto, G., Lemaitre, G., Bactivelane, T. , Wang, M. , Ferrari, M., Mazzanti, S. , Di Biagio, B., & Borra E. F. 1995, A&A Suppl.114 379-386 .

Ninane, N. M., & Jamar C. A. 1996, Applied Optics 35, No 31, 6131.

Ragazzoni, R., & Marchetti, E., 1994 A&A 283, L17 .

Sica, R. J., Sargoytchev, S. , Borra, E.F., Girard, L., Argall, S., Sarrow, C. T., & Flatt, S. 1995, Appl. Opt., 34, No 30, 6925.

Sica, R.J. & Thorsley, M.D. 1996 Geophysical Research Letters Vol. 30, No 20, 2797.